\documentclass[aps,
prb,
preprint,
superscriptaddress,
longbibliography]{revtex4-2}

\usepackage{graphicx}
\usepackage{multirow}
\usepackage{amsthm,amsmath,amsfonts}
\usepackage{comment}
\usepackage{hyperref}
\usepackage{longtable,tabularx}
\usepackage{subfigure,xcolor,color}
\usepackage{siunitx}
\usepackage[utf8]{inputenc}
\usepackage[ruled,vlined]{algorithm2e}
\usepackage{graphicx}
\usepackage{dcolumn}
\usepackage{bbm,url}
\usepackage{tabularx}
\usepackage{bm}
\usepackage{amssymb}
\usepackage{mathtools}

\newcommand{\bsigma}{{\bm\sigma}}

\newcommand{\HH}{\mathcal{H}}
\newcommand{\LL}{\mathcal{L}}
\newcommand{\dt}{\mathrm{d}t}

\newcommand{\eloc}{E_{\mathrm{loc}}}

\begin{document}

\title{Propagating sparsely supported states with time-dependent neural quantum states}

\author{Lexin Ding}
\email{leding@ethz.ch}
\affiliation{ETH Z\"urich, Department of Chemistry and Applied Biosciences, Vladimir-Prelog-Weg 2, CH-8093 Z\"urich, Switzerland}

\author{Markus Reiher}
\email{mreiher@ethz.ch}
\affiliation{ETH Z\"urich, Department of Chemistry and Applied Biosciences, Vladimir-Prelog-Weg 2, CH-8093 Z\"urich, Switzerland}

\begin{abstract}
    Neural quantum states (NQSs) have emerged as a powerful ansatz for quantum dynamics. Their high entanglement capacity promises to overcome the entanglement barrier. 
    However, existing time-dependent NQS methods are ill-equipped to propagate sparsely supported initial states due to the inherent mismatch between the supports of such states and their time derivatives. 
    To address this issue, we introduce the interpolation sampling method 
    as a novel form of importance sampling,
    where samples are drawn from a distribution that interpolates between the wave function and its time derivative.
    Combined with a global-in-time variational principle, we demonstrate with the example of the two-dimensional transverse-field Ising model that interpolation sampling allows for accurate propagation of extremely peaked states that standard wave function sampling could not tackle.
    We further improve sampling efficiency by introducing (i) a configuration-time joint sampling scheme where spin configurations and time are both treated as random variables, and (ii) a hybrid strategy for sample proposal that incorporates the knowledge of a Krylov subspace.
    Our work extends the scope of time-dependent NQS methods to a wider range of physically relevant scenarios, while challenging the prevailing reliance on the Born distribution and its close variants as the default basis for sampling.
\end{abstract}

\maketitle

\section{Introduction}

Simulating real-time dynamics is a major challenge in many-body physics and quantum chemistry. Not only does the Hilbert space dimension grow exponentially with the system size, but the entanglement in the time-evolved quantum state also grows rapidly in time, demanding more and more computational resources to faithfully represent a state. Tensor network (TN) methods allowed for tremendous progress in simulating the dynamics of local systems in one dimension (see Ref.~\cite{paeckel2019time} for a review), with time-evolving block decimation\cite{tebd1,tebd2,tebd3} and time-dependent variational principle\cite{tdvp,haegeman2016unifying} (TDVP) being the two most widely used approaches. However, the local nature of the TN states calls for an exceedingly large bond dimensions to overcome the entanglement barrier. Meanwhile, neural quantum states (NQSs) have emerged as a promising contender for time-dependent simulations\cite{carleo2014light,2017-CarleoTroyer,schmitt2020quantum,Gutierrez2022realtimeevolution,donatella2023dynamics,medvidovi2023variational,nys2024ab}. They are formally free of the locality constraints of TNs, with tremendous flexibility in network architecture ranging from simple restricted Boltzmann machines to highly expressive transformers\cite{glasser2018neural,luo2018backflow,hibat2020recurrent,sharir2020deep,schmale2022arcnn,2022-NAQS,zhang2023transformer,lange2025transformer,denis2025bosoncnn,ibarra2025auto}. Hence, NQSs offer an efficient encoding of highly entangled many-body wave functions\cite{deng2017entanglement, gao2017efficient,levine2019quantum, zakari2025comment}, even those beyond the scope of TNs\cite{sharir2022neural,rpj5-cns6}.

That said, a major limitation of NQS-based methods is that they assume the wave function to have broad support for Monte Carlo sampling to be feasible. 
The most common approach to propagating NQSs, following TDVP, is time-dependent variational Monte Carlo\cite{2017-CarleoTroyer,schmitt2020quantum} (tVMC).
In tVMC, the network parameters are updated in time steps in the direction of the stochastically estimated gradient, involving inverting the so-called quantum geometric tensor, which can be highly singular and requires sophisticated regularization techniques\cite{schmitt2020quantum,hofmann2022role}. Most severely, the parameter update breaks down completely when the wave function support does not fully contain the support of its gradient with respect to the network parameters, and the dynamics freezes\cite{Sinibaldi2023unbiasingtime}. This issue of support mismatch is avoided by projected variants of tVMC where the exact time-evolved state is projected back onto the variational manifold instead of evolving along it\cite{Sinibaldi2023unbiasingtime,nys2024ab,Gravina2025neuralprojected}. However, tVMC remains a sequential method, which makes it prone to accumulation of errors over time.

To avoid accumulation of error, Refs.~\cite{vandewalle2025tnqs,sinibaldi2026galerkin} proposed a novel alternative to tVMC, where the wave function is not sequentially optimized in time, but the trajectory over an entire time interval is optimized to satisfy the time-dependent Schr\"odinger equation. Despite using different NQS ans\"atze, both works minimize the deviation from the time-dependent Schr\"odinger equation through Monte Carlo sampling according to the Born distribution of the wave function at different time points. We therefore refer to this class of approach as global-in-time variational Monte Carlo (gVMC). Similarly to tVMC, gVMC also relies on the positivity of the Born distribution in order to estimate the time derivative in the loss function accurately, making it unsuitable to treat sparsely supported quantum states. Many common initial states in quench dynamics studies such as N\'eel states\cite{cramer2008exploring,wouters2014quench,vincenzo2017entanglement} and domain wall states\cite{santos2011domain,misguich2017dynamics,medenjak2020domain}, or antiferromagnetically ordered states in cold atom experiments\cite{mazurenko2017cold}, do not satisfy this positivity requirement. The key difference from tVMC is that the support mismatch is now between the wave function and its time derivative, which calls for a different solution.

The goal of this work is to address the issue of support mismatch and develop a robust method for propagating sparsely supported initial states with NQSs.
To this end, we propose an interpolation sampling scheme that fundamentally resolves the issue of support mismatch for gVMC. 
In addition, we develop two new sampling strategies that significantly reduce the cost of sampling. In the time domain, we perform Monte Carlo sampling over a continuous interval in time, removing any bias introduced by the discretization in time. 
For sampling in the Hilbert space, we introduce a hybrid strategy for sample proposal that incorporates a bias towards the most important configurations for the early time dynamics, while retaining the ability to explore the rest of the Hilbert space. Together, they form a configuration-time joint sampling method uniquely suited for optimizing a time-dependent NQS over a period of time.

The article is organized as follows: In Section \ref{sec:theory}, we recap the theoretical framework for optimizing the time-dependent neural quantum state via minimizing the loss function, and introduce our new sampling target. In Section \ref{sec:sampling}, we describe the continuous time sampling and hybrid strategy for sample proposal. In Section \ref{sec:results}, we present a numerical demonstration of our method with the transverse-field Ising model (TFIM) in two dimensions, showcasing the improved accuracy and efficiency of our choice of sampling target over alternatives.

\section{Theory} \label{sec:theory}

\subsection{Global-in-time variational Monte Carlo}
We briefly recall the global-in-time variational principle\cite{mclachlan1964variational} and its usage for time evolution in the framework of variational Monte Carlo.
Let $\Psi_{\bm\theta}(\bsigma,t)$ be a time-dependent wave function in the Hilbert space $\HH$ where $\bsigma$ is the vector of physical variables (e.g., spins) and $t$ is time, with the boundary condition ${\Psi_{\bm\theta}(\bsigma,t\!=\!0) \!=\! \Psi_0(\bsigma)}$ where $\Psi_0$ is the time-independent initial state and $\bm{\theta}$ is a vector of parameters. Let $\hat{H}$ be the Hamiltonian that drives the dynamics. $\Psi_{\bm\theta}(\bsigma,t)$ faithfully represents the dynamics of the system in the time interval $[0,T]$ if and only if it satisfies the time-dependent Schr\"odinger equation
\begin{equation}
    \dot{\Psi}_{\bm{\theta}}(\bsigma,t) + i (\hat{H}\Psi_{\bm{\theta}})(\bsigma,t) = 0, \quad \forall\: \bsigma \in \HH, \: t \in [0,T],
\end{equation}
where $\dot{\Psi}_{\bm\theta}=\partial\Psi_{\bm{\theta}}/\partial t$ is the time derivative of the wave function.
Equivalently, the optimal vector of parameters $\bm\theta^\ast$ is a minimizer to the integral-sum of the residual of the time-dependent Schr\"odinger equation
\begin{equation}
    \LL(\bm{\theta}) = \int_0^T \sum_{\bsigma \in \HH} \left|\dot{\Psi}_{\bm{\theta}}(\bsigma,t) + i (\hat{H}\Psi_{\bm{\theta}})(\bsigma,t)\right|^2 \dt , \label{eqn:loss}
\end{equation}
with ${\LL(\bm{\theta}^\ast) \!=\! 0}$ being the minimum. This variational condition is known as the McLachlan variational principle\cite{mclachlan1964variational}. Despite the original claim by McLachlan, it is equivalent to the TDVP behind tVMC and the Dirac-Frenkel variational principle under commonly satisfied conditions\cite{broeckhove1988equivalence}.

In Refs.~\cite{vandewalle2025tnqs,sinibaldi2026galerkin}, the loss function $\LL$ is estimated using Monte Carlo techniques
\begin{equation}
\begin{split}
    \quad\: \LL(\bm{\theta}) &= \int_0^T \sum_{\bsigma \in \HH} |\Psi_{\bm\theta}(\bsigma,t)|^2 \left|\frac{\dot{\Psi}_{\bm\theta}(\bsigma,t)}{\Psi_{\bm\theta}(\bsigma,t)} + i \eloc(\bsigma,t)\right|^2\! \dt
    \\
    &\approx \sum_{t_i}  \mathbb{E}_{\bsigma \sim |\Psi_{\bm\theta}(\bsigma,t_i)|^2 }\left[\left|\frac{\dot{\Psi}_{\bm\theta}(\bsigma,t_i)}{\Psi_{\bm\theta}(\bsigma,t_i)} + i \eloc(\bsigma,t_i)\right|^2\right]\! \Delta t, 
\end{split} \label{eqn:loss_discrete}
\end{equation}
where
\begin{equation}
    \eloc (\bsigma,t) = \frac{(\hat{H}\Psi_{\bm\theta})(\bsigma,t)}{\Psi_{\bm\theta}(\bsigma,t)}
\end{equation}
is the local energy of $\Psi_{\bm\theta}$ at time $t$. Here, the continuous time integral is approximated by a discrete sum over a set of time points $\{t_i\}$, separated by time step $\Delta t$. This global-in-time variational Monte Carlo (gVMC) method sets itself apart from the traditional tVMC in that the wave function is not propagated sequentially in time, but rather optimized as a trajectory over a period of time. This is not to say that the gVMC method is entirely free from error accumulation, as in practice the total time interval for which the dynamics is solved is often broken into smaller segments. However, the lengths of these time segments remain much larger than the size of a time step in the sequential formalism.

We note that, depending on the parameterization of the NQS, the loss function Eq.~\eqref{eqn:loss} might be trivially minimized by the vanishing norm of the wave function. To avoid this, one can use a projection to ensure norm and phase invariance\cite{sinibaldi2026galerkin}. Alternatively, the NQS ansatz can include a static initial state with a fixed norm, which naturally enforces the norm of the optimized wave function at later times\cite{vandewalle2025tnqs}. Our work adopts the latter route.

\subsection{Sharply peaked initial states}

The Monte Carlo estimation of the loss function $\LL$ in Eq.~\eqref{eqn:loss_discrete} relies on the positivity of the time-specific Born distribution, defined as 
\begin{eqnarray}
    P_t(\bsigma) = \frac{|\Psi_{\bm\theta}(\bsigma,t)|^2}{\sum_{\bsigma'}|\Psi_{\bm\theta}(\bsigma',t)|^2}.
\end{eqnarray}
That is, the necessary condition for the Monte Carlo sampling to correctly reflect the explicit sum is that $\Psi_{\bm{\theta}}$ should have support at least as large as that of $\dot{\Psi}_{\bm\theta}$ and $\hat{H}\Psi_{\bm\theta}$, i.e.,
\begin{equation}
    \Psi_{\bm\theta}(\bsigma,t) = 0 \quad \Rightarrow \quad \dot{\Psi}_{\bm\theta}(\bsigma,t) = (\hat{H}\Psi_{\bm\theta})(\bsigma,t) = 0.
\end{equation}
This condition excludes common sparsely supported states such as the fully polarized spin state in the $z$ direction and a N\'eel state on a Hubbard lattice. Mathematically speaking, a finitely supported initial state will lead to a division by zero in Eq.~\eqref{eqn:loss_discrete}. Practically speaking, even though such an initial state may only be approximately zero outside of the support, the Monte Carlo estimation of the loss function will be divergently large. These divergent behaviors will also appear in the Monte Carlo estimation of the loss function gradient and lead to an unstable optimization, as we demonstrate numerically in Section \ref{sec:results}. 

It is worth pointing out that tVMC also struggles with sparsely supported initial states, although through a different mechanism\cite{Sinibaldi2023unbiasingtime}. For gVMC, the support mismatch is between the wave function and its time derivative, whereas for tVMC, the support mismatch occurs between the wave function and its gradients with respect to network parameters. We provide a mathematical argument here for the extreme case where the initial state is only supported on a single configuration $\bsigma_0$. In the tVMC framework, the network parameters $\theta_m$ advance in time according to the equation of motion that solves for the first-order time derivative
\begin{equation}
    \bm{S} \dot{\bm{\theta}} = -i \bm{G},
\end{equation}
where $\bm{S}$ is the quantum geometric tensor (QGT) and $\bm{G}$ is the force vector, which can be expressed as covariances
\begin{equation}
\begin{split}
    S_{mn} &= \langle O_m^\ast O_n \rangle - \langle O_m^\ast \rangle \langle O_n \rangle = \mathrm{Cov}(O_m^\ast,O_n),
    \\
    G_m &= \langle O_m^\ast E_{\rm loc} \rangle - \langle O_m^\ast \rangle \langle E_{\rm loc} \rangle = \mathrm{Cov}(O_m^\ast,E_{\rm loc}),
\end{split}
\end{equation}
where
\begin{equation}
    O_m(\bsigma) = \frac{\partial}{\partial \theta_m} \log(\Psi_{\bm{\theta}}(\bsigma)).
\end{equation}
The QGT encodes geometric information on the variation of the wave function in the state manifold as the network parameters change. It plays a central role in NQS methods, appearing in stochastic reconfiguration\cite{2017-CarleoTroyer,chen2024empowering}, natural gradient descent\cite{Stokes2020quantumnatural}, and time evolution\cite{schmitt2020quantum}.
Note that the covariance of any two functions $g_1$ and $g_2$ with respect to a delta distribution ${P(\bsigma) \!=\! \delta_{\bsigma,\bsigma_0}}$ is always zero, i.e., 
\begin{equation}
\begin{split}
    \mathrm{Cov}(g_1,g_2) &= \mathbb{E}_{\bsigma \sim P}[g_1 g_2] - \mathbb{E}_{\bsigma \sim P}[g_1] \mathbb{E}_{\bsigma \sim P}[g_2] 
    \\
    &= g_1(\bsigma_0) g_2(\bsigma_0) - g_1(\bsigma_0) g_2(\bsigma_0) 
    \\
    &= 0.
\end{split}
\end{equation}
Therefore, both the QGT and the force vector are zero for such an initial state, and the equation of motion is ill-defined, with any $\dot{\bm{\theta}}$ being a solution. In practice, the QGT is inverted with a small regularization term $\lambda \mathbbm{1}$ added to it, which leads to only one solution for the time derivative of the network parameters being ${\dot{\bm{\theta}}\!=\!0}$, essentially freezing the network parameters at their initial values for all subsequent time. This frozen dynamics was observed in Ref.~\cite{Sinibaldi2023unbiasingtime} when the state was driven to a single configuration state. There, the issue of support mismatch was not directly resolved but rather circumvented by an evolve-and-project sequential method, which has no direct analogy in the global-in-time framework. 

Recently, Ref.\cite{wan2026removing} revisited the issue of support mismatch in tVMC and resolved it using a blurred kernel method as a postprocessing step after sampling, which effectively smears the zero nodes in the Born distribution. In the next section, we provide a solution in the context of gVMC directly at the stage of sampling, where the underlying Born distribution is replaced by an interpolation function.

\subsection{Interpolation sampling} \label{subsec:interpolation}

To directly resolve the issue of support mismatch between the wave function and its time derivative in the gVMC framework, we propose a new sampling method, dubbed interpolation sampling.
To this end, the usual reweighting strategies in importance sampling\cite{misery2026looking} would not be useful, and a drastic change of the sampling target is needed.
We first introduce a wave function ansatz as a superposition between two neural networks, one of which is a hardwired initial state $\Psi_0$, and the other is a time-dependent neural network $F_{\bm{\theta}}$ to be optimized
\begin{equation}
    \Psi_{\bm\theta}(\bsigma,t) = (1-f_{\alpha}(t)) \Psi_0(\bsigma) + f_{\alpha}(t) F_{\bm\theta}(\bsigma,t). \label{eqn:interpolation}
\end{equation}
Here, $f_{\alpha}(t)$ is a differentiable function satisfying $f_\alpha(0)\!=\!0$ and $f_{\alpha}(T)\!=\!1$. Ref.~\cite{vandewalle2025tnqs} employed a similar interpolation ansatz, but on the level of the amplitude and phase of the wave function rather than the wave function itself. We find that the direct superposition in Eq.~\eqref{eqn:interpolation} allows for a clearer physical interpretation of $F_{\bm \theta}$. We choose the following rational function for $f_{\alpha}(t)$ 
\begin{equation} \label{eqn:rational}
    f_{\alpha}(t) = \frac{\alpha t}{T + (\alpha-1)t},
\end{equation}
where $\alpha$ is a hyperparameter controlling the rate at which the total wave function transitions from the initial state to the time-dependent network. When ${\alpha\!=\!1}$ the interpolation function is simply $f_1(t)\!=\!t/T$.

\begin{figure}
    \centering
    \includegraphics[width=0.6\textwidth]{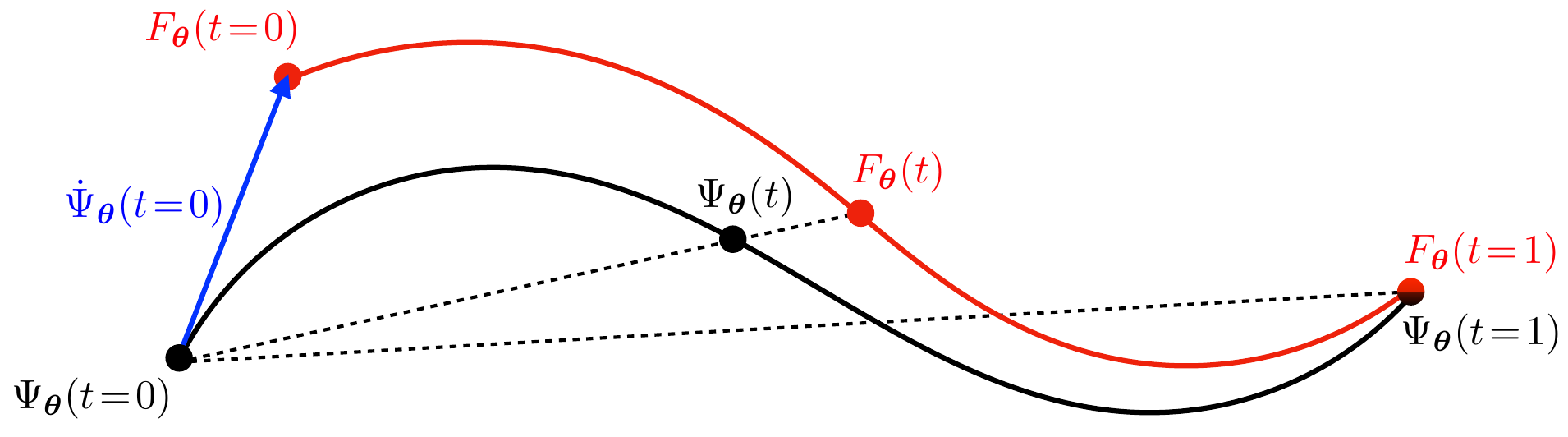}
    \caption{Illustration of the interpolation sampling method in the time interval $[0,1]$. The solid black curve represents the time-dependent wave function $\Psi_{\bm\theta}(t)$, and the solid red curve represents the interpolation function $F_{\bm\theta}(t)$ in Eq.~\eqref{eqn:interpolation} with the scaling function $f_1(t)\!=\!t/T$. The blue solid arrow at $t\!=\!0$ represents the time derivative $\dot{\Psi}(t\!=\!0)$ tangent to $\Psi_{\bm\theta}(t)$.}
    \label{fig:interpolation_sampling}
\end{figure}

The time-dependent neural network $F_{\bm\theta}(\bsigma,t)$ is our new sampling target that does not suffer from the issue of support mismatch. To see this, we can compute the value of $F_{\bm\theta}(\bsigma,t)$ at $t\!=\!0$, where the issue of support mismatch is the most severe. We have
\begin{equation}
    F_{\bm\theta}(\bsigma,0) = \Psi_0(\bsigma) + \dot{\Psi}_{\bm\theta}(\bsigma,0) \frac{T}{\alpha}.
\end{equation}
That is, $F_{\bm\theta}(\bsigma,0)$ is the first-order Taylor expansion of the time-evolved wave function $\Psi_{\bm\theta}(\bsigma,t)$ around ${t\!=\!0}$ evaluated at ${t\!=\!T/\alpha}$. Through this relation, $F_{\bm\theta}(\bsigma,0)$ contains information about the time derivative of the wave function $\dot{\Psi}_{\bm \theta}(\bsigma,0)$ that is crucially missing in the initial state $\Psi_0(\bsigma)$ and that has support at least as large as that of $\dot{\Psi}_{\bm\theta}(\bsigma,0)$ and $\hat{H}\Psi_{\bm\theta}(\bsigma,0)$. We illustrate this property of $F_{\bm\theta}(\bsigma,0)$ for the simplified case of ${\alpha\!=\!1}$ ($f_1(t)\!=\!t/T$) and ${T\!=\!1}$ in Figure \ref{fig:interpolation_sampling}. 
At intermediate time, $F_{\bm\theta}(\bsigma,t)$ is a point on the extended part of the secant line that connects the initial state $\Psi_0(\bsigma)$ and the time-evolved wave function $\Psi_{\bm\theta}(\bsigma,t)$.

In addition to the geometric interpretation, the interpolation function $F_{\bm\theta}(\bsigma,t)$ can also be decomposed into contributions from the time-evolved wave function and its derivatives
\begin{equation}
    F_{\bm\theta}(\bsigma,t) = \Psi_{\bm\theta}(\bsigma,t) + \frac{1-f_{\alpha}(t)}{f_{\alpha}(t)} \sum_{k=1}^{\infty} \frac{t^k}{k!} \frac{\partial^k \Psi_{\bm\theta}(\bsigma,0)}{\partial t^k}.
\end{equation}
Finally, at $t=T$ the interpolation function coincides exactly with the time-evolved wave function, $F_{\bm\theta}(\bsigma,T)=\Psi_{\bm\theta}(\bsigma,T)$.

With the interpolation function $F_{\bm\theta}(\bsigma,t)$ as the sampling target, we can rewrite the loss function for gVMC as
\begin{equation}
\begin{split}
    \LL({\bm\theta}) &= \int_0^T \sum_{\bsigma \in \HH} |F_{\bm\theta}(\bsigma,t)|^2 \left|\frac{\dot{\Psi}_{\bm\theta}(\bsigma,t)}{F_{\bm\theta}(\bsigma,t)} + i \frac{(\hat{H}\Psi_{\bm\theta})(\bsigma,t)}{F_{\bm\theta}(\bsigma,t)}\right|^2 \dt 
    \\
    &= \int_0^T \sum_{\bsigma \in \HH} |F_{\bm\theta}(\bsigma,t)|^2 |\LL_{\rm loc}(\bsigma,t)|^2 \dt, \label{eqn:loss_F}
\end{split}
\end{equation}
where
\begin{eqnarray}
    \LL_{\rm loc}(\bsigma,t) &=& \frac{\dot{\Psi}_{\bm\theta}(\bsigma,t)}{F_{\bm\theta}(\bsigma,t)} + i \frac{(\hat{H}\Psi_{\bm\theta})(\bsigma,t)}{F_{\bm\theta}(\bsigma,t)}
\end{eqnarray}
is the local loss function.
Since we established that $F_{\bm\theta}(\bsigma,t)$ has support at least as large as that of $\dot{\Psi}_{\bm\theta}(\bsigma,t)$ and $\hat{H}\Psi_{\bm\theta}(\bsigma,t)$, the catastrophic division by zero is avoided, and the Monte Carlo estimation of the loss function is well-defined and stable. 

\begin{figure}
    \centering
    \includegraphics[width=0.6\textwidth]{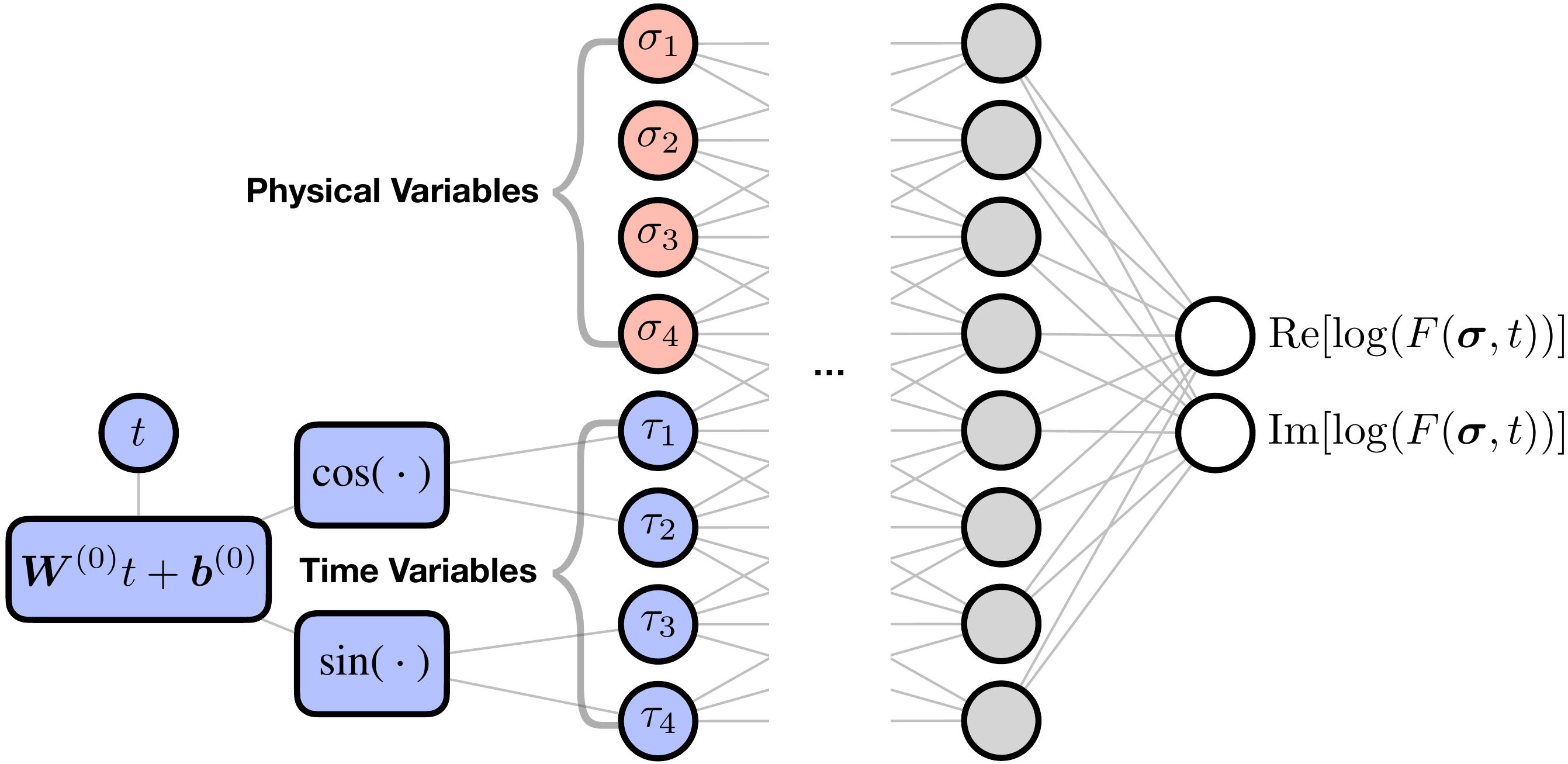}
    \caption{Illustration of the time-dependent neural quantum state (tNQS) architecture for the interpolation function $F_{\bm\theta}(\bsigma,t)$ in Eq.~\eqref{eqn:tau}. The input layer consists of the physical variables $\bsigma$ and transformed time variables $\bm\tau$, which are processed through multiple hidden layers to produce the real and imaginary parts of $\log(F_{\bm\theta}(\bsigma,t))$.}
    \label{fig:schematic}
\end{figure}

\subsection{Time-dependent neural quantum state (tNQS)}

Minimizing the loss function Eq.~\eqref{eqn:loss_F} over a time interval requires an ansatz that represents the quantum state on this entire time interval. We can do so by constructing a time-dependent neural quantum state (tNQS). Neural networks with an explicit time input have been used to solve partial differential equations for the last few decades\cite{lagaris1998artificial,sirignano2018dgm,cuomo2022scientific}. Recently, several tNQS proposals were applied to spin lattice systems\cite{wang2021spacetimeneuralnetworkhigh,vandewalle2025tnqs}, many-electron systems in real space\cite{hou2026global}, and quantum emitters\cite{qccj-6vyt}.  

We choose to encode the time variable $t$ into the neural network through a SIREN subnetwork\cite{sitzmann2019siren}, which is designed to capture periodic features of the input data. 
We transform $t$ via a linear layer followed by sinusoidal activation functions 
\begin{equation}
    \bm{\tau} = (\sin(\bm{W}^{(0)}t + \bm{b}^{(0)}), \cos(\bm{W}^{(0)}t + \bm{b}^{(0)})). \label{eqn:tau}
\end{equation}
Here, the length of the vector $\bm{W}^{(0)}$ formally represents the number of independent frequencies learned by the network, which we set to be the same as the system size $N$ in all our implementations.
In contrast to the original implementation of SIREN, we use both sine and cosine functions to activate the transformed time variables.
The output of the SIREN subnetwork $\bm{\tau}$ is then concatenated with the physical input $\bsigma$ (e.g., spin configuration) to form a composite input variable $\bm{x}^{(0)}$, and passed through a standard multilayer perceptron\cite{rumelhart1986learning} of depth $L$ to produce the final output $F_{\bm\theta}(\bsigma,t)$
\begin{equation}
    \begin{split}
        &\bm{x}^{(0)} = (\bsigma, \bm{\tau}),
        \\
        &\bm{x}^{(l)} = \tanh(\bm{W}^{(l)}\bm{x}^{(l-1)} + \bm{b}^{(l)}), \quad l=1,\ldots,L-1,
        \\
        &\bm{x}^{(L)} = \bm{W}^{(L)}\bm{x}^{(L-1)} + \bm{b}^{(L)} Eq.~\equiv (x_0^{(L)}, x_1^{(L)}),
        \\
        &F_{\bm\theta}(\bsigma,t) = \exp\left(x^{(L)}_0 + i x^{(L)}_1\right),
    \end{split}
\end{equation}
where the weight matrices $\bm{W}^{(l)}$ and bias vectors $\bm{b}^{(l)}$ are collected into the parameter vector $\bm\theta$.
The network architecture is summarized in Figure \ref{fig:schematic}. In later sections, we use ${N_{\rm hidden}\!=\![d_1, d_2, \ldots d_{L-1}]}$ to denote the widths of the hidden layers.

The network parameters $\theta_m$ are updated using the loss function gradients
\begin{equation}
    \begin{split}
        \frac{\partial \LL}{\partial \theta_m} &=  \frac{\partial}{\partial \theta_m} \int_0^T \sum_{\bsigma \in \HH} (|F_{\bm\theta}(\bsigma,t)|^2 |\LL_{\rm loc}|^2) \dt
        \\
        &=\int_0^T \sum_{\bsigma \in \HH} |F_{\bm\theta}(\bsigma,t)|^2 |\LL_{\rm loc}(\bsigma,t)|^2
        \\
        & \quad \quad  \times \frac{\partial}{\partial\theta_m} \log(F_{\bm\theta}(\bsigma,t)\LL_{\rm loc}(\bsigma,t)) \dt + {\rm {\rm H.c.}} 
    \end{split} \label{eqn:gradients}
\end{equation}
Similarly to the loss function, the integral-sum in the gradients is estimated using Monte Carlo sampling, which we will discuss in the next section.

\section{Monte Carlo sampling} \label{sec:sampling}

\subsection{Configuration-time joint sampling}

Previously, the time integral in the loss function Eq.~\eqref{eqn:loss} was approximated by a discrete sum over a set of time points $\{t_i\}$\cite{vandewalle2025tnqs,sinibaldi2026galerkin}, introducing a bias in the training. This bias can be reduced by increasing the number of points in the time interval, but this comes with an increased computational cost.
Conceptually, time-local sampling strategies do not fully exploit the representational power of the neural network over the continuous time segment.
Instead, we can treat the time variable $t$ as a continuous random variable, and perform a joint sampling directly in the composite space $\mathcal{H}\times[0,T]$. Sampling from the time domain has been done in the study of neural network solutions for partial differential equations\cite{sirignano2018dgm}, but not yet in NQS studies.
Specifically, we use the following Metropolis-Hastings algorithm\cite{hastings1970monte} to form a Markov chain of configuration-time samples $(\bsigma,t)$:

\begin{enumerate}
    \item Initialize the chain with a configuration-time sample $(\bsigma,t)$.
    \item Propose a time point $t'$ uniformly from the interval $[0,T]$.
    \item Propose a spin configuration $\bsigma'$ with a transition probability ${P(\bsigma \!\to\! \bsigma')}$ based on the current configuration $\bsigma$.
    \item Accept or reject the proposed configuration-time sample $(\bsigma',t')$ according to the Metropolis-Hastings criterion. That is, the new configuration-time sample $(\bsigma',t')$ is accepted with probability
    \begin{equation}
        A((\bsigma,t) \to (\bsigma',t')) = \min\left(1, \frac{|F_{\bm\theta}(\bsigma',t')|^2}{|F_{\bm\theta}(\bsigma,t)|^2}\right).
    \end{equation}
    \item Update the current configuration-time sample $(\bsigma,t)$ to the new sample $(\bsigma',t')$ if accepted, and repeat from step 2 until the desired number of samples is reached.
\end{enumerate}

Once the set of samples $\{(\bsigma_i,t_i)\}_{i=1}^{N_s}$ of size $N_s$ is generated, the loss function in Eq.~\eqref{eqn:loss_F} can be estimated as
\begin{equation}
\begin{split}
    \frac{\LL(\Psi)}{Z} &= \mathbb{E}_{(\bsigma,t)\sim|F_{\bm\theta}(\bsigma,t)|^2} \left[|\LL_{\rm loc}(\bsigma_i,t_i)|^2\right]
    \\
    &\approx \frac{1}{N_s} \sum_{i=1}^{N_s} |\LL_{\rm loc}(\bsigma_i,t_i)|^2, \label{eqn:loss_vmc}
\end{split}
\end{equation}
and similarly for the gradients in Eq.~\eqref{eqn:gradients}
\begin{equation}
\begin{split}
    \frac{1}{Z}\frac{\partial \LL}{\partial \theta_m} 
    &= \mathbb{E}_{(\bsigma,t)\sim|F_{\bm\theta}(\bsigma,t)|^2}  \left[|\LL_{\rm loc}(\bsigma_i,t_i)|^2 \phantom{\frac{\partial}{\partial\theta_m}}\right.
    \\
    & \quad  \times \left.\frac{\partial}{\partial\theta_m} \log(F_{\bm\theta}(\bsigma_i,t_i)\LL_{\rm loc}(\bsigma_i,t_i)) + {\rm H.c.}\right],
    \\
    &\approx \frac{1}{N_s} \sum_{i=1}^{N_s} \left[ |\LL_{\rm loc}(\bsigma_i,t_i)|^2 \phantom{\frac{\partial}{\partial\theta_m}} \right.
    \\
    & \quad  \times \left. \frac{\partial}{\partial\theta_m} \log(F_{\bm\theta}(\bsigma_i,t_i)\LL_{\rm loc}(\bsigma_i,t_i)) + {\rm H.c.}\right],
\end{split}
\end{equation}
where $Z=\int_0^T \sum_{\bsigma \in \mathcal{H}} |F_{\bm\theta}(\bsigma,t)|^2 \, \mathrm{d}t$ is the normalization prefactor.

\subsection{Hybrid proposal in the Hilbert space} \label{subsec:hybrid}

A major challenge we face when sampling a sparsely supported state is that the relevant region of the Hilbert space can be exponentially small, and the Markov chain takes a long time to reach this region by random exploration. In practice, the inability to efficiently explore the relevant regions of the Hilbert space results in a low acceptance rate during sampling, an increase in the computational cost, and an inaccurate estimation of the underlying distribution\cite{choo2020fermionic}. To address this issue, we propose a hybrid strategy for sample proposal that combines knowledge of the relevant subspace with the ability to explore beyond it.

We first generate a Krylov set by taking the following union
\begin{equation}
    \mathcal{S}_K = \bigcup_{k=0}^{K} \hat{H}^k \mathcal{S}_0,
\end{equation}
where ${\mathcal{S}_0\!=\!\{\bsigma\!\in\!\HH|\Psi_0(\bsigma)\!\neq\!0\}}$ is the set of configurations for which the initial wave function is nonzero. The order $K$ is a hyperparameter that controls the size of the Krylov set. Note that the Krylov set is a basis of the Krylov subspace\cite{parlett1998symmetric}, which is the linear span of $\{|\psi_0\rangle, \hat{H}|\psi_0\rangle, \ldots, \hat{H}^K|\psi_0\rangle\}$ for a given state $|\psi_0\rangle$, and plays an essential role in various time-dependent methods\cite{manmana2005time,yang2020krylov,takahashi2025krylov}. The construction of the Krylov set is closely related to the Taylor expansion of the time evolution operator $e^{-i\hat{H}t}$
\begin{equation}
    e^{-i\hat{H}t} = \sum_{k=0}^{\infty} \frac{(-i\hat{H}t)^k}{k!},
\end{equation}
which at early time can be truncated to a finite number of terms while maintaining a good approximation of the exact dynamics entirely within the span of a finite-order Krylov set.

The hybrid strategy for sample proposal combines two proposal rules, one that draws configurations from the Krylov set ${\mathcal{S}_K}$ (the Krylov rule), and the local rule that randomly flips one spin in the current configuration. A new configuration is proposed with the Krylov rule with probability $p$, and with the local rule with probability ${1\!-\!p}$. The parameter $p$ is a hyperparameter that controls the degree of bias towards the Krylov set. In this work, we set ${p\!=\!0.5}$ as a default. In this way, the Markov chain is given direct access to the relevant region of the Hilbert space by the Krylov rule, while still retaining the ability to explore the remaining part of the Hilbert space via standard exploration rules. 

It should be noted that the hybrid nature of this proposal strategy breaks detailed balance\cite{fichthorn1991theoretical}, which requires the transition probability to satisfy ${P(\bsigma\!\to\!\bsigma')\!=\!P(\bsigma'\!\to\!\bsigma)}$ for any two configurations. To see this, let us suppose that $\bsigma \in \mathcal{S}_K$ and $\bsigma' \notin \mathcal{S}_K$, such that $\bsigma'$ differs from $\bsigma$ by two spin flips. The transition probability from $\bsigma$ to $\bsigma'$ is zero, since neither proposal rule can propose $\bsigma'$ from $\bsigma$. However, the transition probability from $\bsigma'$ to $\bsigma$ is ${p/|\mathcal{S}_K|}$ where $|\mathcal{S}_K|$ is the cardinality of the Krylov set.
The asymmetry in the proposal kernel breaks the detailed balance condition and results in a bias in the sampled distribution. That said, this bias is not necessarily harmful to the training, since the Krylov set in fact captures the most relevant region of the Hilbert space for the targeted dynamics. Moreover, the local loss function is driven to zero during training, which means that small deviations in the actual distribution from the exact distribution are well tolerated. In practice, we find that the hybrid proposal technique significantly improves the training efficiency for propagating sparsely supported states by saving time for the Markov chain to explore the exponentially large Hilbert space to reach the relevant configurations.

\section{Results} \label{sec:results}

\subsection{Computational methodology}

For NQS implementations in this work, we use the libraries NetKet\cite{NetKet,NetKet3}, JAX\cite{JAX}, and Flax\cite{Flax}, whereas matrix product state (MPS) calculations for reference are performed with the Python library TeNPy \cite{tenpy2024}. MPSs of bond dimension $64$ and $128$ are propagated with a time step of 0.001 using the TDVP method. The neural networks are trained with the continuous resilient (CoRe) optimizer\cite{core1,core2} with an exponentially decaying learning rate $\eta = \eta_0\, r^{m/M}$ where $r$ is the decay rate, $m$ is the index of the current optimization step, and $M$ is the transition length of the decay. The default values for the hyperparameters are $\eta_0 = 0.005$, $r=0.5$, $M=1000$. To ensure a fair comparison, we use the same network architecture for all three sampling modes for a given initial state. We also use an explicit summation method instead of Monte Carlo to compute the observables of interest, which is still feasible for the $4\!\times\!4$ lattice with a Hilbert space of dimension $2^{16}$. This eliminates any errors occurring from the measurement, allowing us to focus on the errors arising from optimization of the time evolution only.

\subsection{Transverse-field Ising model}

In this section, we demonstrate the effectiveness of our method by propagating sparsely supported initial states with the Transverse-field Ising model (TFIM) Hamiltonian on a two-dimensional square lattice.
The TFIM Hamiltonian is given by
\begin{equation}
    \hat{H}_{\rm TFIM} = -\sum_{\langle i,j\rangle} \hat{\sigma}_i^z \hat{\sigma}_j^z - h \sum_i \hat{\sigma}_i^x.
\end{equation}
This system is critical at $h\!=\!h_c\!=\!3.04438$ in two dimensions \cite{tfim_critical}.
We consider the dynamics starting from three initial states with different supports in the computational basis in the $z$ direction:

\begin{enumerate}
    \item \textit{$z$-polarized state ${|\!\uparrow\,\rangle^{\otimes N}}$.} This state has support on a single configuration and is essentially a delta function in the Hilbert space. The spin alignment along the $z$ direction in this state is characterized by the average magnetization
    \begin{equation}
        \overline{\langle \sigma^z \rangle} = \frac{1}{N} \sum_{\bm{r}} \langle \hat{\sigma}_{\bm{r}}^z \rangle,
    \end{equation}
    where $\bm{r} = (i,j)$ is the coordinate of the spin on the two-dimensional lattice.  This state can be conveniently represented by a delta neural network with the following mapping
    \begin{eqnarray}
        \langle \bsigma|\!\uparrow\rangle^{\otimes N} \propto \prod_{i=1}^N \delta_{\sigma_i, \uparrow} .
    \end{eqnarray}

    \item \textit{Single magnon state}\cite{auerbach2012interacting}
    \begin{equation}
        |k_x,k_y\rangle = \frac{1}{\sqrt{N}} \sum_{\bm{r}} e^{i \bm{k}\cdot\bm{r}} \hat{S}^-_{\bm{r}} |\!\uparrow\uparrow\cdots\uparrow\rangle.
    \end{equation}
    This state has support on all single-spin-flip configurations. The flipped spin is delocalized over the entire lattice with a wave vector $\bm{k} = (k_x,k_y)$. The number of magnons in this momentum state is measured by the magnon number operator
    \begin{equation}
        \hat{n}_{\bm{k}} = \frac{1}{N} \sum_{\bm{r},\bm{r}'} e^{i \bm{k}\cdot(\bm{r}-\bm{r}')} \hat{S}^-_{\bm{r}} \hat{S}^+_{\bm{r}'}.
    \end{equation}
    Magnons are bosonic quasiparticles that represent the collective excitations of the spin system. Therefore, the value of the magnon number $\langle \hat{n}_{\bm{k}} \rangle$ can be larger than 1. We focus on the state with $k_x=0$ and $k_y=0$, which can be straightforwardly represented by the following network mapping
    \begin{eqnarray}
        \langle \bsigma | k_x\!=\!0,k_y\!=\!0\rangle \propto \exp\left[-R\left|1 - N + \sum_{i=1}^N \delta_{\sigma_i,\uparrow}\right|\right],
    \end{eqnarray}
    where $R$ is a large exponent such that the coefficients of the configurations other than a single spin-flip are exponentially suppressed. In our implementation, we set $R=30$.

    \item \textit{$x$-polarized state}
    \begin{equation}
    \begin{split}
        |\!\rightarrow\rangle^{\otimes N} &= \frac{1}{2^{N/2}} \bigotimes_{i=1}^N (|\!\uparrow\rangle_i + |\!\downarrow\rangle_i)
        \\
        &= \frac{1}{2^{N/2}} \sum_{\bsigma \in \HH} |\bsigma\rangle.
    \end{split}
    \end{equation}
    This state is a uniform superposition of all configurations in the Hilbert space. Its spin alignment along the $x$ direction is characterized by the average magnetization in the $x$ direction
    \begin{equation}
        \overline{\langle \sigma^x \rangle} = \frac{1}{N} \sum_{\bm{r}} \langle \hat{\sigma}_{\bm{r}}^x \rangle.
    \end{equation}
    We simulate the $x$-polarized state trivially with the identity network
    \begin{eqnarray}
        \langle \bsigma |\!\rightarrow\rangle^{\otimes N} \propto 1.
    \end{eqnarray}
\end{enumerate}

\begin{figure}
    \centering
    \includegraphics[width=0.6\textwidth]{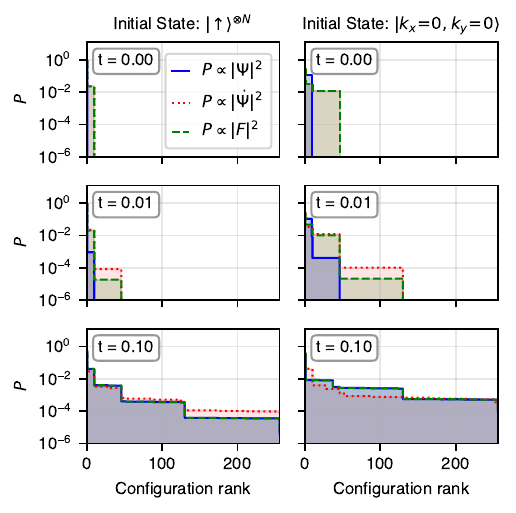}
    \caption{The time-local normalized distributions $P\!\propto\!|\Psi|^2$, $|\dot{\Psi}|^2$, and $|F|^2$ in Eq.~\eqref{eqn:time_local_dist} during the time evolution of the $z$-polarized state ${|\!\uparrow\rangle^{\otimes N}}$ and the single-magnon state ${|k_x\!=\!0,k_y\!=\!0\rangle}$ driven by the $3\!\times\!3$ TFIM Hamiltonian with $h\!=\!h_c$ for $T\!=\!0.1$, evaluated at $t\!=\!0$, 0.01, and 0.1. The scaling function $f_1(t)\!=\!t/T$ is used for the interpolation. Only the 250 configurations with the largest probability coefficients are shown.}
    \label{fig:distribution_comparison}
\end{figure}

We first verify the existence of the problem of support mismatch for the sparsely supported initial states.
In Figure \ref{fig:distribution_comparison}, we show the distribution of the wave function $|\Psi(t)|^2$, its time derivative $|\dot{\Psi}(t)|^2$, and the interpolation function $|F(t)|^2$ normalized in a time local manner, i.e.,
\begin{eqnarray}
    P(\bsigma,t) = \frac{|X(\bsigma,t)|^2}{\sum_{\bsigma \in \HH} |X(\bsigma,t)|^2}, \quad X=\Psi,\dot{\Psi},F, \label{eqn:time_local_dist}
\end{eqnarray}
computed by exact unitary evolution for the $z$-polarized state and the single-magnon state $|k_x\!=\!0,k_y\!=\!0\rangle$  driven by the $3\!\times\!3$ TFIM Hamiltonian for $T\!=\!0.1$. We used the scaling function $f_1(t)\!=\!t/T$ for the interpolation in Eq.~\eqref{eqn:interpolation}.
Only the 250 configurations with the highest probability coefficients are shown.
For the $z$-polarized state, the distribution of $|\Psi|^2$ is only supported on one single configuration at $t\!=\!0$, whereas $|\dot{\Psi}|^2$ is non-zero on exactly nine configurations. By contrast, the interpolation function distribution $|F|^2$ is almost identical to the distribution of the time derivative $|\dot{\Psi}|^2$. At time ${t\!=\!0.1}$, all three distributions start to spread. However, $|F|^2$ still reflects the distribution of the time derivative better than that of the wave function, and there remain regions where $|\dot{\Psi}|^2$ and $|F|^2$ are non-zero, but $|\Psi|^2$ is zero.
This shows that the issue of support mismatch is present not only at ${t\!=\!0}$, but also persists over an extended period at early times.
At the end of the time interval, $|F|^2$ becomes identical to $|\Psi|^2$, as expected. But at this point, all three distributions are broadly supported in the Hilbert space, and the support mismatch disappears. For the initial state $|k_x\!=\!0, k_y\!=\!0\rangle$, the discrepancy between the distributions of $|\Psi|^2$ and $|\dot{\Psi}|^2$ is even more pronounced at ${t\!=\!0}$. At early time, in addition to support mismatch, there are a considerable number of configurations whose $|F|^2$ and $|\dot{\Psi}|^2$ values are almost two orders of magnitude larger than the corresponding $|\Psi|^2$ values. Figure \ref{fig:distribution_comparison} demonstrates the issue of support mismatch between the wave function and its time derivative at early time of the evolution, and the ability of $|F|^2$ to exactly bridge this gap. 

\begin{figure}
\centering
\includegraphics[scale=0.9]{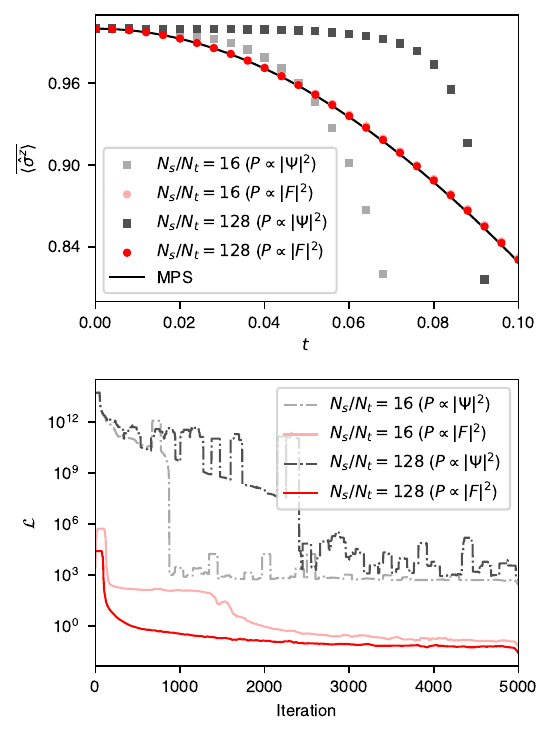}
\caption{Average magnetization $\overline{\langle \sigma^z \rangle}$ over time and the loss function $\LL$ during the optimization of the dynamics of the $z$-polarized state ${|\!\uparrow\rangle^{\otimes N}}$ driven by the $4\!\times\!4$ TFIM Hamiltonian with $h\!=\!h_c$ for $T\!=\!0.1$, trained with state sampling ($P\!\propto\!|\Psi|^2$) and interpolation sampling ($P\!\propto\!|F|^2$). The sampling of the Hilbert space is performed separately at $N_t=21$ discrete time points evenly spaced on the time interval with number of samples per time point $N_s/N_t=16,128$, and without hybrid proposal. Both sampling modes use the same network architecture, with $N_{\rm hidden}=[32,32]$.}
\label{fig:discrete}
\end{figure}

Due to the ability of the interpolation function $F$ to capture the distribution of the time derivative $\dot{\Psi}$ at early times, the improvement brought about by interpolation sampling to the gVMC formalism is dramatic. We isolate this improvement by performing a comparison between state sampling (${P\!\propto\!|\Psi|^2}$) and interpolation sampling (${P\!\propto\!|F|^2}$) for the $z$-polarized state driven by the $4\!\times\!4$ TFIM Hamiltonian, without configuration-time joint sampling and hybrid proposal. That is, we assign a fixed number of samples $N_s/N_t$ for each of the evenly spaced $N_t=21$ time points $t_i$ within the time interval of $T\!=\!0.1$, including end points, and the integral of time in the loss function is estimated with a discrete sum according to Eq.~\eqref{eqn:loss_discrete} and similarly to Refs.~\cite{vandewalle2025tnqs,sinibaldi2026galerkin}. In Figure \ref{fig:discrete}, we present the time evolution of the average magnetization $\overline{\langle \sigma^z \rangle}$ and the loss function $\LL$ (averaged over a window of 100 iterations) for both sampling modes and sample sizes per time point $N_s/N_t=16,128$. We observe that overall the dynamics obtained with state sampling significantly deviates from the MPS calculations. The time-evolved state freezes at the initial state for the first half of the time interval, indicated by the flattened magnetization curve. After that, the average magnetization rapidly decreases. The loss function $\LL$ in this sampling mode is highly noisy and does not converge to a small value. Notably, the performance of the state sampling mode does not improve with increasing sample size $N_s/N_t$. This is because increasing the sample size only makes the problem of support mismatch more evident, rather than resolving it. By contrast, the optimized dynamics from interpolation sampling closely follows the reference MPS calculations. Although the loss function first hits a plateau for the case $N_s/N_t=16$, the optimization eventually improves and converges to the correct dynamics at later iterations. The plateau is completely removed with a higher number of samples, indicating a systematic improvability with the sample size. Together, Figures \ref{fig:distribution_comparison} and \ref{fig:discrete} demonstrate that the problem of support mismatch is detrimental to the propagation of sparsely supported initial states and that interpolation sampling offers a direct solution to it.

\begin{figure*}
    \centering
    \includegraphics[width=0.9\textwidth]{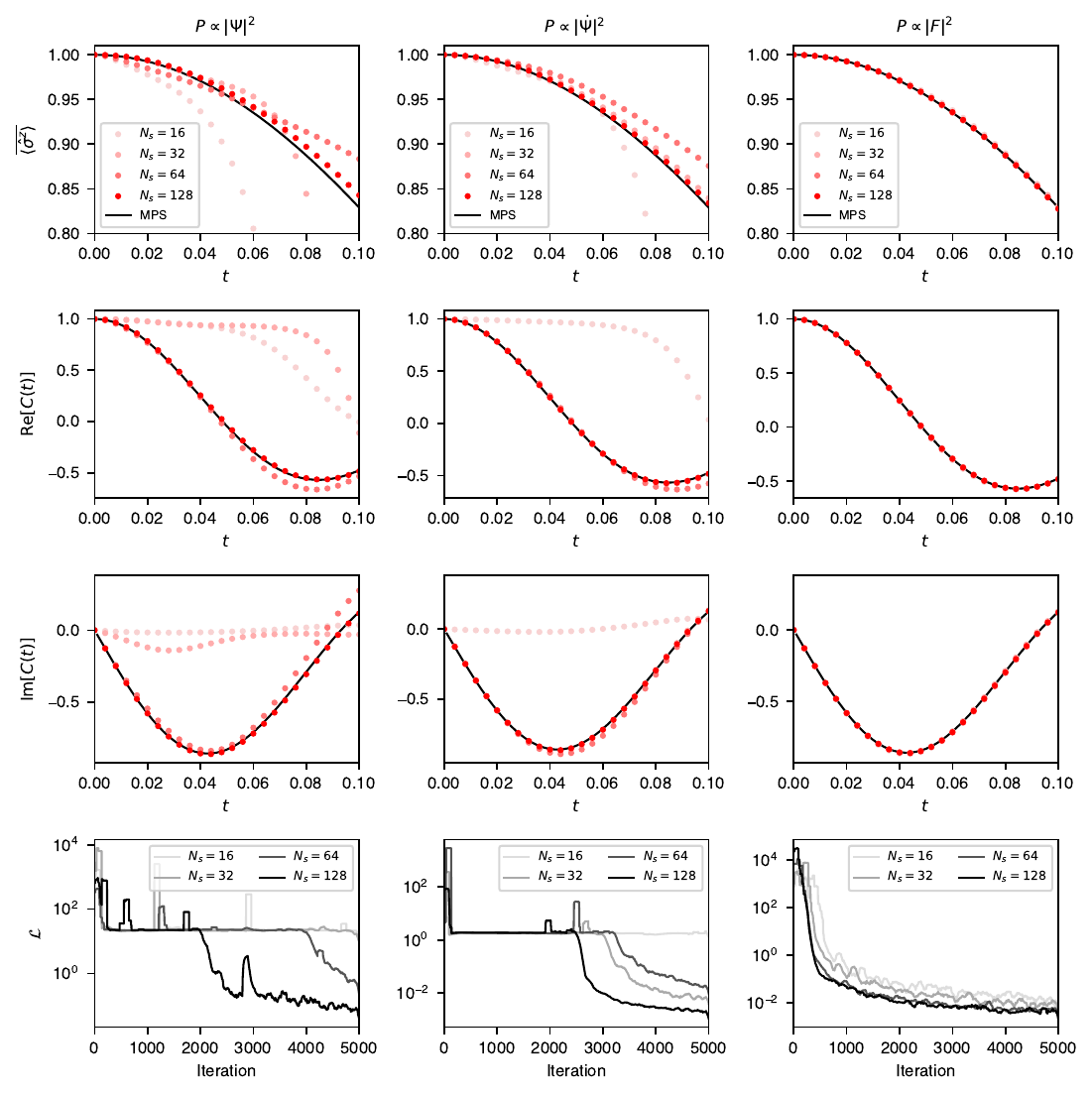}
    \caption{Average magnetization $\overline{\langle \sigma^z\rangle}$ and autocorrelation function $C(t)$ over time, and the loss function $\LL$ for the dynamics of the $z$-polarized state ${|\!\uparrow\rangle^{\otimes N}}$ under the $4\!\times\!4$ TFIM Hamiltonian with $h\!=\!h_c$ for $T\!=\!0.1$, trained with state sampling ($P\!\propto\!|\Psi|^2$), derivative sampling ($P\!\propto\! |\dot{\Psi}|^2$), and interpolation sampling ($P\!\propto\!|F|^2$), in combination with $N_s\!=\!16,32,64,128$ configuration-time samples. Configuration-time joint sampling and hybrid proposal rule with $K\!=\!4$ are used. All calculations use the same network architecture for the interpolation function $F$, with $N_{\rm hidden}=[32,32]$.}
    \label{fig:N16_coh0.0}
\end{figure*}

Next, we apply the configuration-time joint sampling and hybrid proposal techniques to further improve the efficiency of the optimization. Figure \ref{fig:N16_coh0.0} presents a side-by-side comparison of dynamics obtained with three sampling modes: (i) state sampling ($P \!\propto\! |\Psi|^2$), (ii) derivative sampling ($P \!\propto\! |\dot{\Psi}|^2$), and (iii) interpolation sampling ($P \!\propto\! |F|^2$). The derivation for the loss function in the derivative sampling mode is similar to that in Section \ref{subsec:interpolation}, with $\dot{\Psi}$ replacing $F$ in all instances.
We simulate the dynamics starting from the $z$-polarized state with the $4\!\times\!4$ TFIM Hamiltonian for $T\!=\!0.1$. All three sampling modes use the same network architecture with $N_{\rm hidden}=[32,32]$, and the optimization is performed with 16, 32, 64, and 128 configuration-time samples. We set the order $K$ of the Krylov set for the hybrid proposal to be $K=4$. After the optimization, we computed the average magnetization $\overline{\langle\sigma^z\rangle}$, as well as the real and imaginary parts of the autocorrelation function 
\begin{eqnarray}
C(t) = \langle \Psi(t) | \Psi(0) \rangle = \langle \Psi(0) | e^{i\hat{H}t}|\Psi(0)\rangle.  
\end{eqnarray}
Finally, we compare the loss function (averaged over a window of 100 iterations). Note that since the normalization factor of the three distributions
\begin{equation}
    Z_X = \int_0^T \sum_{\bsigma\in\HH} |X(\bsigma,t)|^2\,\mathrm{d}t, \quad X = \Psi, \, \dot{\Psi}, \, F
\end{equation}
are different, the absolute value of the loss function should only be compared within one sampling mode but not across different ones.

The results of the state sampling mode deviate most significantly from the MPS calculations. From the autocorrelation function, we can see that for $N_s=16$ and $N_s=32$ the time-evolved states stay approximately the same as the initial state for a prolonged period of time, indicated by $C(t)\approx1$. The failure to learn the correct dynamics by this sampling mode is also reflected by the long plateau of the loss function at low sample sizes. 
These issues are eventually overcome with a sufficiently large number of samples, indicating that configuration-time joint sampling with continuous time can be systematically improved by increasing the sample size, even in the state sampling mode. 
We recall from Figure \ref{fig:discrete} that in the discrete time setting, the time point $t\!=\!0$ where the support mismatch is the most severe is always included in the estimation of the loss function and its gradient, leading to a failed minimization. By contrast, in the configuration-time joint sampling scheme, configuration-time samples at exactly ${t\!=\!0}$ are almost never sampled (since ${t\!=\!0}$ has zero measure in the interval $[0,T]$), thereby alleviating the problem of the support mismatch to some extent.

Surprisingly, the derivative sampling mode also fails to learn the correct dynamics at low sample sizes. At ${N_s\!=\!16}$ it performs similarly to the state sampling mode, where the time-evolved state remains close to the initial state for a long period of time. Moreover, the optimization of the loss function also shows prolonged plateaus that are only remedied by increasing the sample size. However, once the plateaus are resolved, the optimized dynamics quickly converges to the reference results. Overall, the derivative sampling mode is still more effective and efficient than the state sampling mode, as it requires a smaller number of samples to escape the plateau and reach a similar accuracy in the dynamics. 

In stark contrast to the previous two sampling modes, the interpolation sampling mode consistently produces accurate results even for the lowest sample size of ${N_s\!=\!16}$, for both the magnetization and the autocorrelation function. The improvement in the sampling accuracy achieved by the interpolation sampling mode is also reflected in the optimization of the loss function, which shows a steady decrease without any prolonged plateaus, unlike the previous two sampling modes. 
Comparing within the interpolation sampling mode (between Figures \ref{fig:discrete} and \ref{fig:N16_coh0.0}), the implementation of configuration-time joint sampling allows us to produce more accurate results with $N_s=16$ configuration-time samples than the interpolation sampling mode with discretized time and time-local sampling with $N_s=16N_t = 336$ samples. 
Overall, Figure \ref{fig:N16_coh0.0} clearly demonstrates the advantage of interpolation sampling over the other two sampling modes, in terms of both accuracy and efficiency (sampling cost and convergence speed), for propagating the extremely peaked $z$-polarized state. At the same time, configuration-time joint sampling offers further sampling efficiency compared to the discretized time setting.

\begin{figure*}
    \centering
    \includegraphics[width=0.9\textwidth]{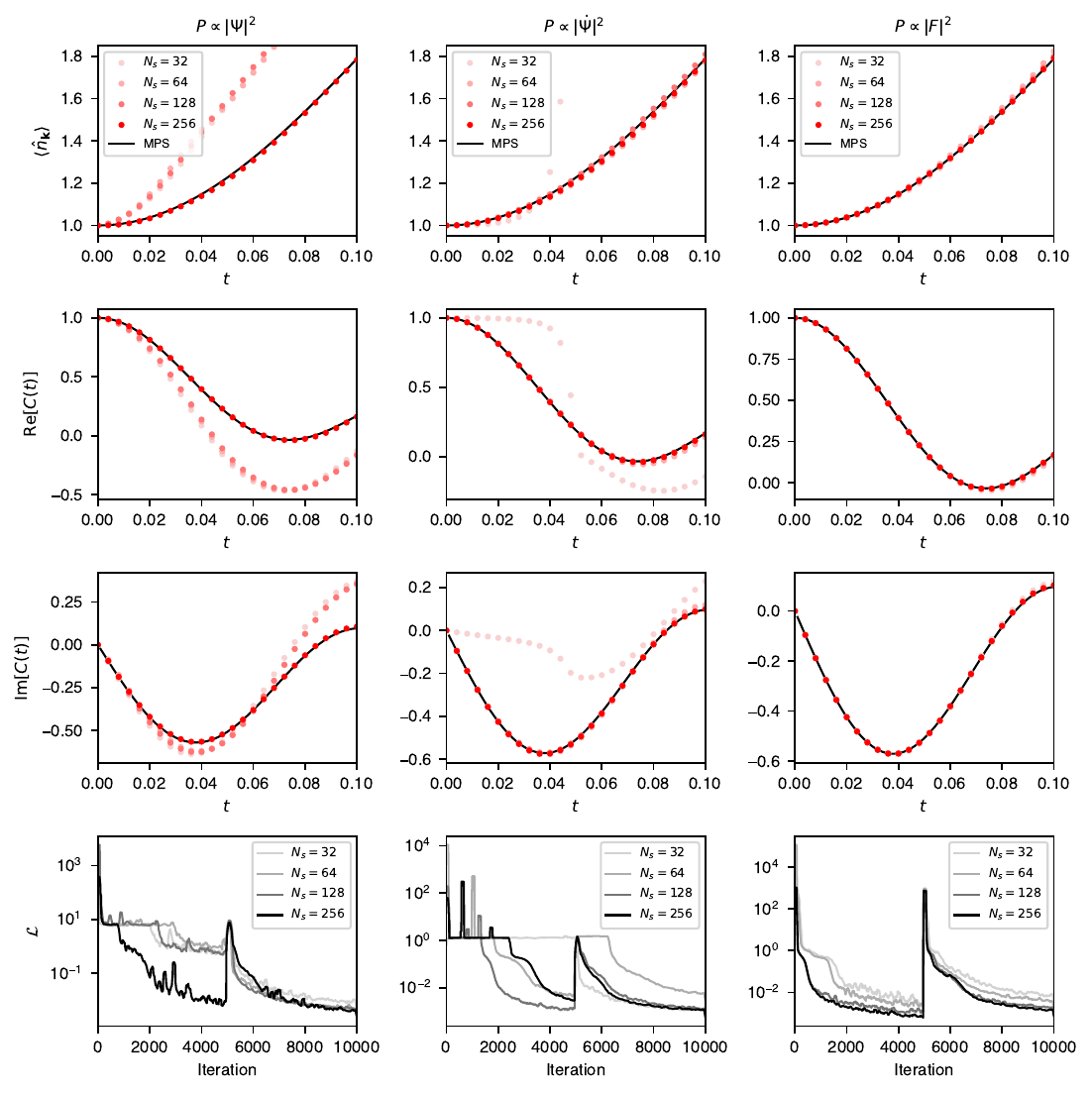}
    \caption{Magnon occupation number $n_{\bm{k}}$ and autocorrelation function $C(t)$ over time, and the loss function $\LL$ for the dynamics of the single-magnon state $|k_x\!=\!0,k_y\!=\!0\rangle$ under the $4\!\times\!4$ TFIM Hamiltonian with $h\!=\!h_c$ for $T\!=\!0.1$, trained with state sampling ($P\!\propto\!|\Psi|^2$), derivative sampling ($P\!\propto\! |\dot{\Psi}|^2$), and interpolation sampling ($P\!\propto\!|F|^2$), in combination with $N_s\!=\!32,64,128,256$ configuration-time samples. Configuration-time joint sampling and hybrid proposal with $K\!=\!4$ are used. All calculations use the same network architecture for the interpolation function $F$, with $N_{\rm hidden}=[48,48]$.}
    \label{fig:N16_kx0.0_ky0.0}
\end{figure*}

We proceed to perform the same comparison of the three sampling modes for the dynamics of a single-magnon state ${|k_x\!=\!0,k_y\!=\!0\rangle}$ driven by the $4\!\times\!4$ TFIM Hamiltonian for $T\!=\!0.1$. The results are shown in Figure \ref{fig:N16_kx0.0_ky0.0}. We used the same network architecture for all three sampling modes with $N_{\rm hidden}=[48, 48]$ and a range of configuration-time sample sizes $\{32,64,128,256\}$. We trained the dynamics for this time interval with two equal length time segments. The results are similar to those for the $z$-polarized state. The results obtained by state sampling remain qualitatively incorrect until $N_s$ reaches 256, while the derivative sampling mode also fails to learn the correct dynamics for $N_s=32$. Even with higher sample sizes, the derivative sampling mode still results in a non-negligible error in the magnon occupation compared to the reference value. Both sampling modes show similar prolonged plateaus that require increasing the sample size to escape. By contrast, the interpolation sampling mode is robust to the low sample sizes and consistently produces accurate results for both the magnon occupation and the autocorrelation function. 

\begin{figure*}
    \centering
    \includegraphics[width=0.9\textwidth]{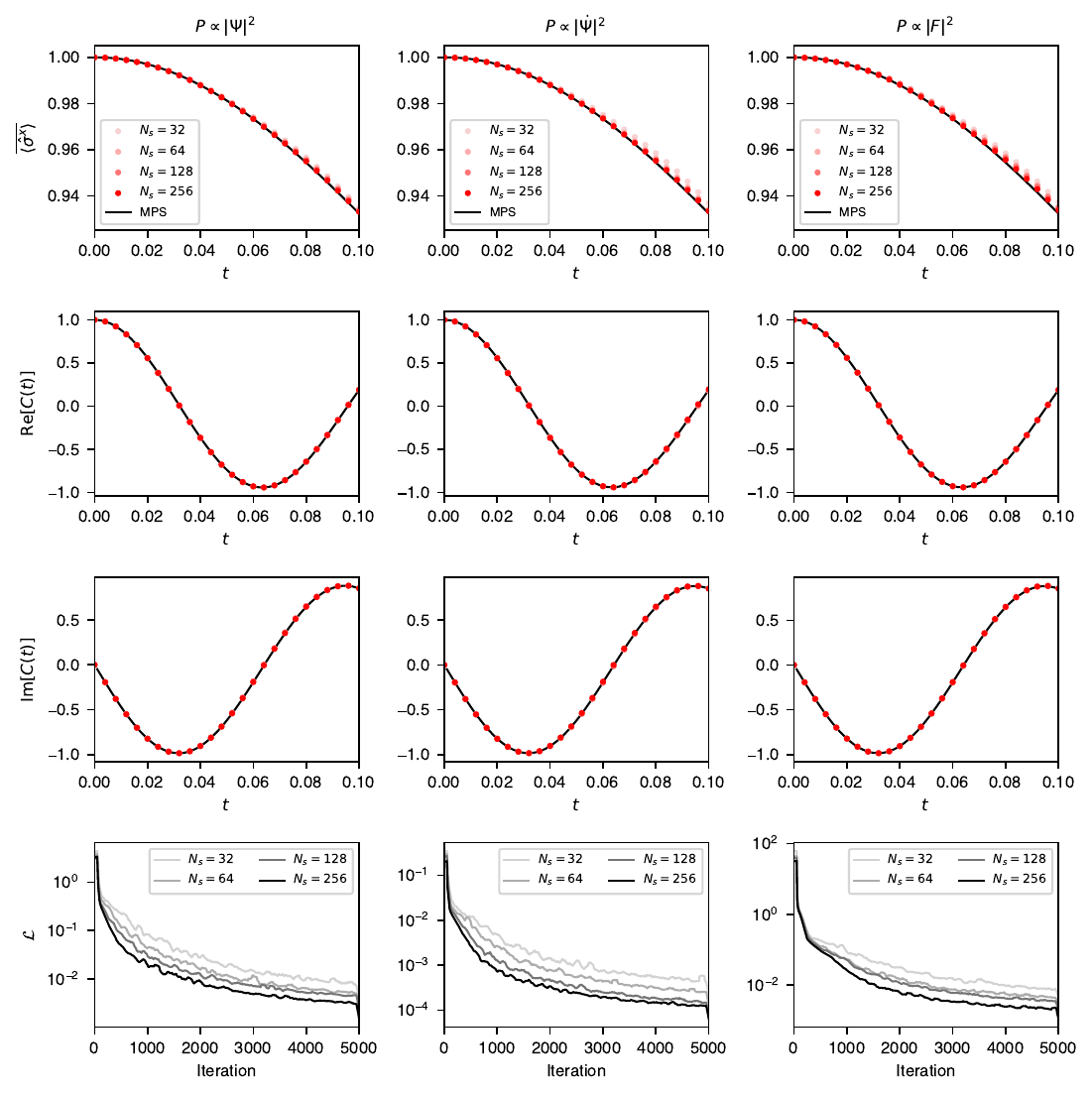}
    \caption{Average magnetization $\overline{\langle \sigma^x\rangle}$ and autocorrelation function $C(t)$ over time, and the loss function $\LL$ for the dynamics of the $x$-polarized state ${|\!\rightarrow\rangle^{\otimes N}}$ under the $4\!\times\!4$ TFIM Hamiltonian with $h\!=\!h_c$ for $T\!=\!0.1$, trained with state sampling ($P\!\propto\!|\Psi|^2$), derivative sampling ($P\!\propto\! |\dot{\Psi}|^2$), and interpolation sampling ($P\!\propto\!|F|^2$), in combination with $N_s\!=\!32,64,128,256$ configuration-time samples. Configuration-time joint sampling and local proposal rule are used. All calculations use the same network architecture for the interpolation function $F$, with $N_{\rm hidden}=[64,64]$.}
    \label{fig:N16_coh0.5}
\end{figure*}

We also investigate the performance of the three sampling modes for the dynamics of the $x$-polarized state. The results are presented in Figure \ref{fig:N16_coh0.5}. This serves as a control experiment, since this initial state is broadly supported in the Hilbert space. We used the same network architecture for all three sampling modes with $N_{\rm hidden}=[64, 64]$ and a range of configuration-time sample sizes $\{32,64,128,256\}$. Only the local proposal rule is used, since the Krylov set would simply be the entire Hilbert space in this case. We see in Figure \ref{fig:N16_coh0.5} that the limitations we observed in the previous two scenarios are no longer present for the state sampling and derivative sampling modes. All three sampling modes now produce similarly accurate results for average magnetization $\overline{\langle\sigma^x\rangle}$ and the autocorrelation function $C(t)$. No plateaus are observed in the optimization of the loss function for any of the three sampling modes, and the convergence speed is improved similarly for all three modes when the sample size is increased. This comparison demonstrates that the problematic behaviors of the state sampling and derivative sampling modes are in fact due to the sparsely supported nature of the initial states, and that the interpolation sampling mode is particularly advantageous for propagating such states.

\begin{figure}[t]
    \centering
    \includegraphics[scale=0.9]{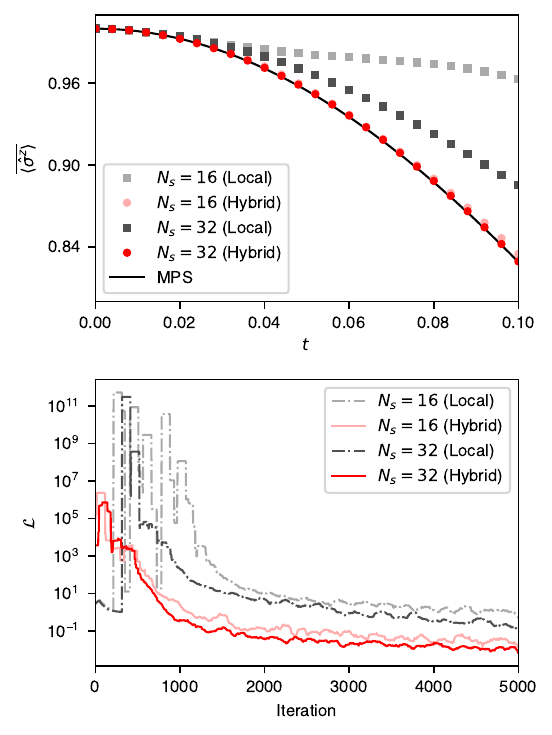}
    \caption{Average magnetization $\overline{\langle \sigma^z \rangle}$ over time and the loss function $\LL$ during the optimization of the dynamics of the $z$-polarized state ${|\!\uparrow\rangle^{\otimes N}}$ driven by the $4\!\times\!4$ TFIM Hamiltonian with $h\!=\!h_c$ for $T\!=\!0.1$, trained with interpolation sampling with local and hybrid proposal rules. Configuration-time joint sampling is used. All calculations use the same network architecture for the interpolation function $F$, with $N_{\rm hidden}=[32,32]$.}
    \label{fig:hybrid_vs_local_comparison}
\end{figure}

For the time propagation of the $z$-polarized and single-magnon states, we used a hybrid proposal method with a Krylov set of order $K=4$. To demonstrate the effectiveness of the hybrid proposal compared to the standard local spin-flip proposal, we perform a comparison with the dynamics of the $z$-polarized state with $N_s=16,32$. The results are shown in Figure \ref{fig:hybrid_vs_local_comparison}. We observe that the optimization of the loss function becomes extremely noisy with divergently large values when the local spin-flip proposal is used. This is because the Markov chains cannot reach the relevant configurations and are stuck with configurations with small network output, leading to large values of the local loss function (recall that the local loss function contains terms that are inversely proportional to the network output). The suboptimal optimization eventually leads to qualitatively inaccurate results for the magnetization. By contrast, the hybrid proposal method ensures a steady minimization of the loss function and produces accurate magnetization results even at low sample sizes, as already demonstrated in Figure \ref{fig:N16_coh0.0}. Figure \ref{fig:hybrid_vs_local_comparison} highlights the critical role of the hybrid proposal method, as an orthogonal aspect to the choice of sampling modes. 

\begin{figure}[t]
    \centering
    \includegraphics[scale=0.9]{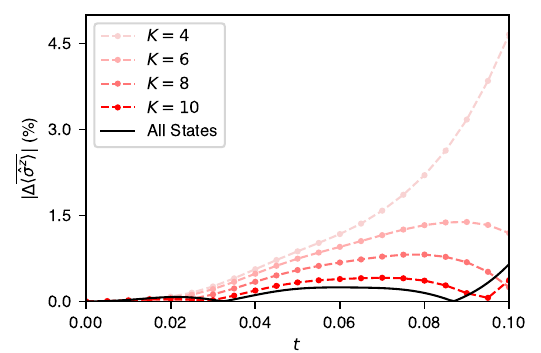}
    \caption{Absolute error $|\Delta\overline{\langle \sigma^z\rangle}|$ of the average magnetization (as a percentage of the range of variation in the time interval $[0,T]$) of the dynamics of the $z$-polarized state ${|\!\uparrow\rangle^{\otimes N}}$ driven by the $4\!\times\!4$ TFIM Hamiltonian with $h\!=\!h_c$ for $T\!=\!0.1$. The average magnetization is measured within the subspace spanned by ${\mathcal{S}_K}$ for $K=4,6,8,10$. A network of size $N_{\rm hidden}=[32,32]$ is used and trained with 64 configuration-time samples using configuration-time sampling and hybrid proposal rule with $K\!=\!4$.}
    \label{fig:n_expand_comparison}
\end{figure}

As discussed in Section \ref{subsec:hybrid}, the hybrid proposal introduces a bias in the sampled distribution that favors the configurations within the Krylov set.
We should therefore eliminate the possibility that the Hilbert space is underexplored, and that the accuracy of the optimized dynamics is achieved only because the wave function does not propagate beyond the Krylov set.
To this end, we measure the average magnetization $\overline{\langle \sigma^z\rangle}$ for the dynamics starting from the $z$-polarized state optimized with $N_s=64$ (taken from Figure \ref{fig:N16_coh0.0}), where the measurement is performed using a selected configuration method restricted to Krylov sets with $K=4,6,8,10$, i.e.
\begin{equation}
\begin{split}
    \overline{\langle\sigma^z\rangle}(t) &\approx \frac{\sum_{\bsigma \in \mathcal{S}_K} |\Psi(\bsigma,t)|^2 
    \sigma^z_{\rm loc}(\bsigma,t) }{\sum_{\bsigma \in \mathcal{S}_K} |\Psi(\bsigma,t)|^2},
    \\
    \sigma^z_{\rm loc} &= \frac{1}{N} \sum_{i=1}^N \frac{(\hat{\sigma}^z_i \Psi)(\bsigma,t)}{\Psi(\bsigma,t)}.
\end{split}
\end{equation}
In Figure \ref{fig:n_expand_comparison} we show the absolute error of the magnetization measurement as a percentage of the range of variation of the reference data within this time interval. We observe that the error decreases from around 5\% to less than 1\% as $K$ increases from 4 to 10 (2517 and 58651 unique configurations, respectively), clearly demonstrating that the network is well-trained for the vast number of relevant configurations outside of the Krylov set of order 4, which is used for the training.

Figure \ref{fig:n_expand_comparison} also shows that to accurately measure the magnetization, it is necessary to include a sufficient number of unique configurations regardless of the measurement method used (VMC or selected configurations). In terms of expansion order, it would require at least $K=8$ to achieve a percentage error below 1\%. This quickly becomes impractical for larger systems, as the number of unique configurations is on the order of $N^K$ for $K\ll N$. Techniques such as importance sampling\cite{misery2026looking} would alleviate the cost of obtaining samples from peaked distributions, and combined with the error cancellation effect from VMC, importance sampling can lead to considerable improvements. However, after testing with overdispersed wave functions proposed in Ref.~\cite{misery2026looking}, we were not able to obtain reliable measurement results with importance sampling for systems larger than $4\!\times\!4$. Therefore, for the larger systems we only measured the autocorrelation function, which can be accurately obtained from a smaller number of unique configurations, as our initial states are sparsely supported.

\begin{figure}[t]
    \centering
    \includegraphics[scale=0.9]{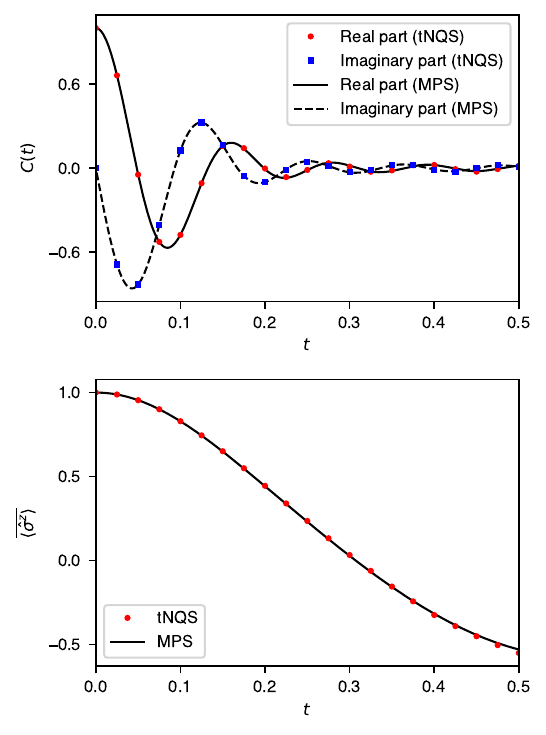}
    \caption{Autocorrelation function $C(t)$ and the average magnetization $\overline{\langle \sigma^z\rangle}$ over time for the dynamics of the $z$-polarized state ${|\!\uparrow\rangle^{\otimes N}}$ driven by the $4\!\times\!4$ TFIM Hamiltonian with $h\!=\!h_c$ for $T=0.5$, trained in five time segments each with a neural network of size $N_{\rm hidden}=[64,64,64]$ using interpolation sampling, configuration-time joint sampling, hybrid proposal rule with $K\!=\!4$, and $N_{s}=512$ configuration-time samples.}
    \label{fig:N16_long_dynamics}
\end{figure}

In Figure \ref{fig:N16_long_dynamics} we show the time evolution of the magnetization and the loss function for the dynamics of the $z$-polarized state under the $4\!\times\!4$ TFIM Hamiltonian with $h\!=\!h_c$ and $T=0.5$. The time interval was divided into five segments of equal length, each represented by a neural network of size $N_{\rm hidden}=[64,64,64]$. Sampling was performed with $N_{s}=512$ configuration-time samples. Each network was trained for $10000$ iterations with a decay transition length $M=2000$ for the learning rate. The reference values for the magnetization and autocorrelation function were obtained from MPS calculations. After the first segment, the wave function was dispersed enough so that the hybrid proposal in the Hilbert space can be replaced by a standard local spin-flip proposal. The autocorrelation function result shows that the perfect spin alignment is quickly lost after the quench, while small revival amplitudes to the initial state are observed at later times. The loss of ferromagnetic order is also reflected in the magnetization, which moves steadily away from the initial value of 1. Overall, the results demonstrate that the piecewise training technique can reliably propagate the NQS in time while maintaining high accuracy compared to the reference results obtained with MPS.

\begin{figure}[t]
    \centering
    \includegraphics[scale=0.9]{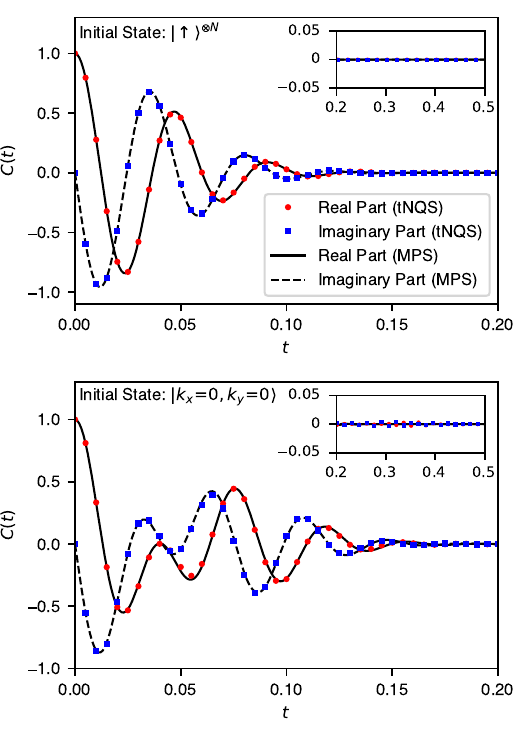}
    \caption{Autocorrelation function $C(t)$ over time for the dynamics of the $z$-polarized state ${|\!\uparrow\rangle^{\otimes N}}$ and single-magnon state $|k_x\!=\!0,k_y\!=\!0\rangle$ driven by the $8\!\times\!8$ TFIM Hamiltonian with $h\!=\!h_c$ for $T=0.5$. 10 time segments equal in length are used for the training for the $z$-polarized state and 25 time segments for the single-magnon state. Each segment is trained with a neural network of size $N_{\rm hidden}=[256,256]$ using interpolation sampling, configuration-time joint sampling, hybrid proposal rule with $K\!=\!3$, and $N_{s}=1024$ configuration-time samples.}
    \label{fig:N64_T0.5}
\end{figure}

Finally, we demonstrate the scalability of our method by simulating the dynamics of the $z$-polarized state and a single-magnon state ${|k_x\!=\!0,k_y\!=\!0\rangle}$ driven by the $8\!\times\!8$ TFIM Hamiltonian with $h\!=\!h_c$ for $T=0.5$. 
The time interval was divided into ten and twenty five segments of equal length for the two initial states, respectively, each represented by a neural network of size $N_{\rm hidden}=[256,256]$. ${N_s\!=\!1024}$ configuration-time samples were used for the training. 
Each network was trained for 20000 steps with a learning rate $\eta=0.0005$ and a decay transition length $M=10000$.
We measured the autocorrelation function for both initial states, and the results are shown in Figure \ref{fig:N64_T0.5}, along with reference data obtained from MPS calculations with bond dimension $128$ and time step 0.001 using the TDVP method. 
Compared to the $4\!\times\!4$ system, the Hilbert space dimension of the $8\!\times\!8$ system is increased by a factor of $2^{48}\approx 2.8\times10^{14}$. However, neural networks can be trained accurately with only 1024 configuration-time samples. 
The larger system size leads to a more rapid loss of the ferromagnetic order for the fully polarized state, as reflected in the fast decay of the magnitude of the autocorrelation function. 
Moreover, the larger eigenenergies of the Hamiltonian in the $8\!\times\!8$ system result in a faster oscillation of the autocorrelation function before it decays to zero. 
After ${t\!=\!0.15}$ the autocorrelation function becomes essentially zero and this behavior is accurately captured by the neural networks. 
The autocorrelation function of the single-magnon state is more complex, and remains non-zero slightly longer than that of the fully polarized state.
In general, the dynamics for the magnon state is also accurately captured in both the oscillation period and the decay rate, with small errors of the order of $10^{-3}$ in the regions where $C(t)=0$.

\section{Conclusions}

In this work, we addressed the challenge of propagating sparsely supported states with neural quantum states (NQS) and global-in-time variational Monte Carlo (gVMC). Our contributions are threefold: (i) We identified the core issue as the support mismatch between the wave function and its time derivative, and resolved it by introducing the interpolation function as the basis of sampling. (ii) We implemented a configuration-time joint sampling scheme, where time is treated as a continuous random variable. (iii) To accommodate the highly peaked quantum states, we developed a hybrid strategy for sample proposal which incorporates knowledge of the Krylov subspace without sacrificing the ability of the Markov chains to explore the remaining part of the Hilbert space. 

With the transverse-field Ising model, we first verified the detrimental effect of support mismatch on the original formulation of gVMC where the quantum state is used as the sampling basis. In direct contrast, we then demonstrated through an extensive comparison that the new interpolation sampling method is the key to an accurate estimation of the gradient of the loss function, exactly because of its correct support over all relevant configurations. We further showed that configuration-time joint sampling significantly reduces the required sample size for accurate time propagation and that the hybrid strategy for sample proposal drastically improves the speed of convergence by saving exploration time. Finally, we tested our method on an $8\!\times\!8$ lattice, showcasing the ability of our method to propagate extremely sparsely supported states in high dimensions, with support taking up only $5\!\times\!10^{-18}$\% ($2^{-64}$) of the Hilbert space.

Sparsely supported states are notoriously challenging to treat with NQSs due to their inherent difficulty to sample. At the same time, such states are prototypical and naturally emerging in various fields of many-body physics and quantum chemistry.
Our work significantly extended the scope of time-dependent NQS and the gVMC formalism beyond states with broad support in the Hilbert space, bringing NQS one step closer to being a universally applicable ansatz. In particular, our solution to the problem of support mismatch led to a new form of importance sampling that is not merely a reweighting of the Born distribution but a substantive change of its support, which may have applications beyond the time-dependent problem. In addition, our configuration-time joint sampling method removes the misalignment between the representational power of time-dependent NQSs and conventional time-local sampling strategies. By putting time on an equal footing with configurations, it establishes a more suitable foundation for future advances in NQS solutions for the time-dependent problem.

Several directions are of interest for future work. First, the potential problem of support mismatch for open quantum dynamics should be investigated with sequential\cite{hartmann2019open} or global-in-time\cite{qccj-6vyt} NQS approaches. Second, due to the high connectivity of ab initio Hamiltonians, the problem of support mismatch is expected to become even more severe for electronic systems. Addressing such non-lattice Hamiltonians will therefore be an important challenge for future developments of time-dependent NQS and gVMC methods. Finally, while configuration-time joint sampling alleviates the challenge of sampling highly peaked distributions in the combined configuration-time domain, sampling the corresponding time-local distributions in the Hilbert space remains challenging. Developing tailored importance sampling strategies for these distributions will be essential for accurately extracting physical observables from the resulting dynamics.

\section*{Acknowledgment}
This work was supported by an ETH Postdoctoral Fellowship.

%

\end{document}